\documentclass{elsart}

\usepackage{graphicx} 
\graphicspath{{ps}}
\newcommand{\mpp}{M_{p\bar{p}}}
\newcommand{\mb}{{M_{\rm bc}}}
\newcommand{\de}{{\Delta{E}}}
\newcommand{\bp}{{B^{+}}}
\newcommand{\bz}{{B^{0}}}
\newcommand{\pp}{{p\bar{p}}}
\newcommand{\ppk}{{p\bar{p}K^+}}
\newcommand{\pppi}{{p\bar{p}\pi^+}}

\newcommand{\ppkst}{{p\bar{p}K^{* +}}}

\newcommand{\plpi}{{p\bar{\Lambda}\pi^-}}

\begin{document}
\begin{frontmatter}

\begin{flushright}
                    Belle Preprint 2007-26 \\
                    KEK \ Preprint 2007-15
\end{flushright}

\title{ \quad\\[1cm]\large
 Study of the decay mechanism for $B^+ \to \ppk$ and $\bp \to \pppi$}


\collab{Belle Collaboration}
  \author[Taiwan]{J.-T.~Wei}, 
  \author[Taiwan]{M.-Z.~Wang}, 
  \author[KEK]{I.~Adachi}, 
  \author[Tokyo]{H.~Aihara}, 
  \author[BINP]{V.~Aulchenko}, 
  \author[Lausanne,ITEP]{T.~Aushev}, 
  \author[Sydney]{A.~M.~Bakich}, 
  \author[ITEP]{V.~Balagura}, 
  \author[Melbourne]{E.~Barberio}, 
  \author[Lausanne]{A.~Bay}, 
% ...

\address[BINP]{Budker Institute of Nuclear Physics, Novosibirsk, Russia}
% ...
\address[KEK]{High Energy Accelerator Research Organization (KEK), Tsukuba, Japan}
% ...
\address[ITEP]{Institute for Theoretical and Experimental Physics, Moscow, Russia}
% ...
\address[Lausanne]{Swiss Federal Institute of Technology of Lausanne, EPFL, Lausanne, Switzerland}
% ...
\address[Melbourne]{University of Melbourne, Victoria, Australia}
% ...
\address[Taiwan]{Department of Physics, National Taiwan University, Taipei, Taiwan}
% ...
\address[Sydney]{University of Sydney, Sydney, NSW, Australia}
% ...
\address[Tokyo]{Department of Physics, University of Tokyo, Tokyo, Japan}
% ...

 


\normalsize
\begin{abstract}
We study the characteristics of the low mass $p\bar{p}$ enhancements near
threshold in the three-body decays $\bp \to \ppk$ and $\bp \to \pppi$.
We observe that the proton polar angle distributions
in the $p\bar{p}$ helicity frame in the two decays
have the opposite polarity, and
measure the forward-backward asymmetries as a function of
the $p\bar{p}$ mass for the $\ppk$ mode. 
We also search for the intermediate two-body
decays, $\bp \to \bar{p} \Delta^{++}$ and $\bp \to p {\bar\Delta^0}$,
and set upper limits on their branching fractions.
These results are obtained from a $414\,{\rm fb}^{-1}$ data sample
that contains 449 $ \times 10^6 B\bar{B}$ events 
collected near the $\Upsilon(4S)$ resonance
with the Belle detector at the KEKB asymmetric-energy $e^+ e^-$
collider.

\end{abstract}
\end{frontmatter}

{\renewcommand{\thefootnote}{\fnsymbol{footnote}}
\setcounter{footnote}{0}

After the first observation of the charmless baryonic $B$ meson decay,
$\bp \to \ppk$~\cite{ppk,conjugate}, many three-body baryonic decays were
found~\cite{plpi,pph,LLK,plg}. The dominant contributions for these decays
are presumably via the $b \to s$ penguin diagram, 
shown in Fig.~\ref{fg:pppifeyn} (a); however $\bp \to
\pppi$  is believed to proceed via the $b \to u$ 
tree diagram as shown in Fig.~\ref{fg:pppifeyn} (b). 
One interesting feature 
of these decays is that the dibaryon mass 
spectra show enhancements near threshold as conjectured in Ref.~\cite{HS}. 
Many theoretical explanations~\cite{theory} have been proposed to 
describe these enhancements in the dibaryon system,
which seem to be a universal feature of all charmless
baryonic $B$ decays.
Study  of the proton polar angular distribution for the dibaryon system
in the $\ppk$ mode~\cite{polar} indicates a violation of the
$b \to s$ short distance picture~\cite{HYCheng}.  
Explicit predictions for 
the dibaryon mass spectra~\cite{mass} 
and the angular distributions 
~\cite{geng,MSuzuki} 
for $\bp \to \ppk/\pi^+$ became available
after the experimental findings were reported.

In this paper, we study the three-body charmless baryonic $B$ meson decays
$\bp \to \ppk$ and  $\bp \to \pppi$.
The differential 
branching fractions as a function of the dibaryon mass and  
the polar angle distributions of 
the proton in the dibaryon system are presented. We also search for
intermediate two-body decays in $\pppi$ three-body final states. This is 
motivated by the observations of two-body decays of charmed 
baryons~\cite{Lcp}. Many predictions based on 
QCD sum rules~\cite{QCDSUM},
pole models~\cite{POLE} and a topological approach~\cite{chua} indicate 
that $\bp \to \bar{p} \Delta^{++}$ and  $\bp \to p {\bar\Delta^0}$
should be observable in the large data samples 
accumulated at the B-factories.   

We use a  414 fb$^{-1}$  data sample,
corresponding to 449 $ \times 10^6 B\bar{B}$ pairs,
collected with the Belle detector 
at the KEKB asymmetric-energy $e^+e^-$ (3.5 on 8~GeV) collider~\cite{KEKB}.
The Belle detector is a large-solid-angle magnetic
spectrometer that
consists of a silicon vertex detector (SVD),
a 50-layer central drift chamber (CDC), an array of
aerogel threshold Cherenkov counters (ACC),
a barrel-like arrangement of time-of-flight
scintillation counters (TOF), and an electromagnetic calorimeter
composed of CsI(Tl) crystals located inside
a super-conducting solenoid coil that provides a 1.5~T
magnetic field.  An iron flux-return located outside of
the coil is instrumented to detect $K_L^0$ mesons and to identify
muons.  The detector
is described in detail elsewhere~\cite{Belle}.
Two inner detector configurations were used. A 2.0 cm beampipe
and a 3-layer silicon vertex detector were used for the first sample
of 152 $\times 10^6 B\bar{B}$ pairs, while a 1.5 cm beampipe, a 4-layer
silicon detector and a small-cell inner drift chamber were used to record
the remaining 297 $\times 10^6 B\bar{B}$ pairs~\cite{Ushiroda}.

\vskip 1cm
\begin{figure}[htb]
\begin{center}
\hskip -3cm {\bf (a)} \hskip 5.5cm {\bf (b)}\\
\vskip -0.8cm
\includegraphics[width=0.44\textwidth]{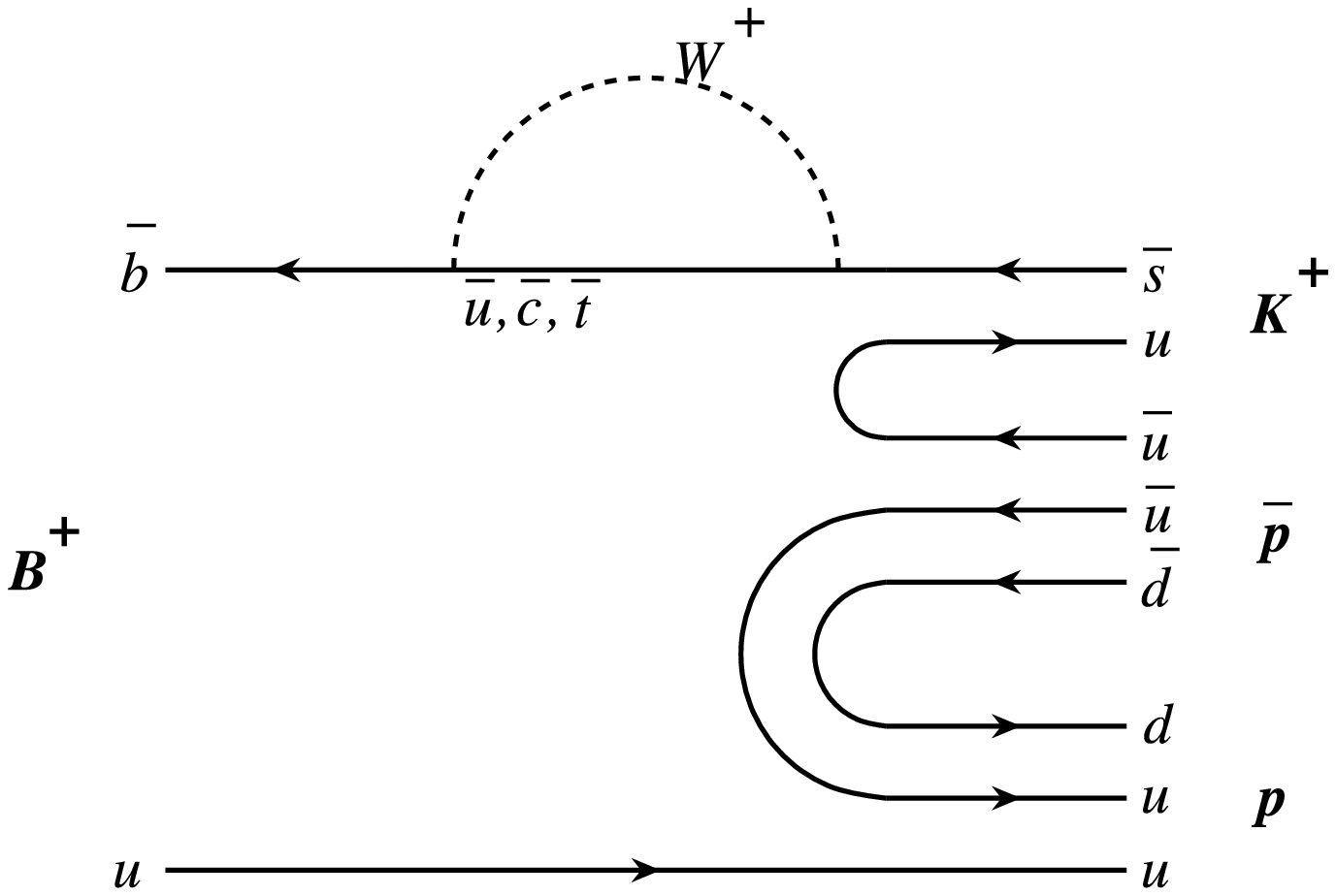}
\includegraphics[width=0.44\textwidth]{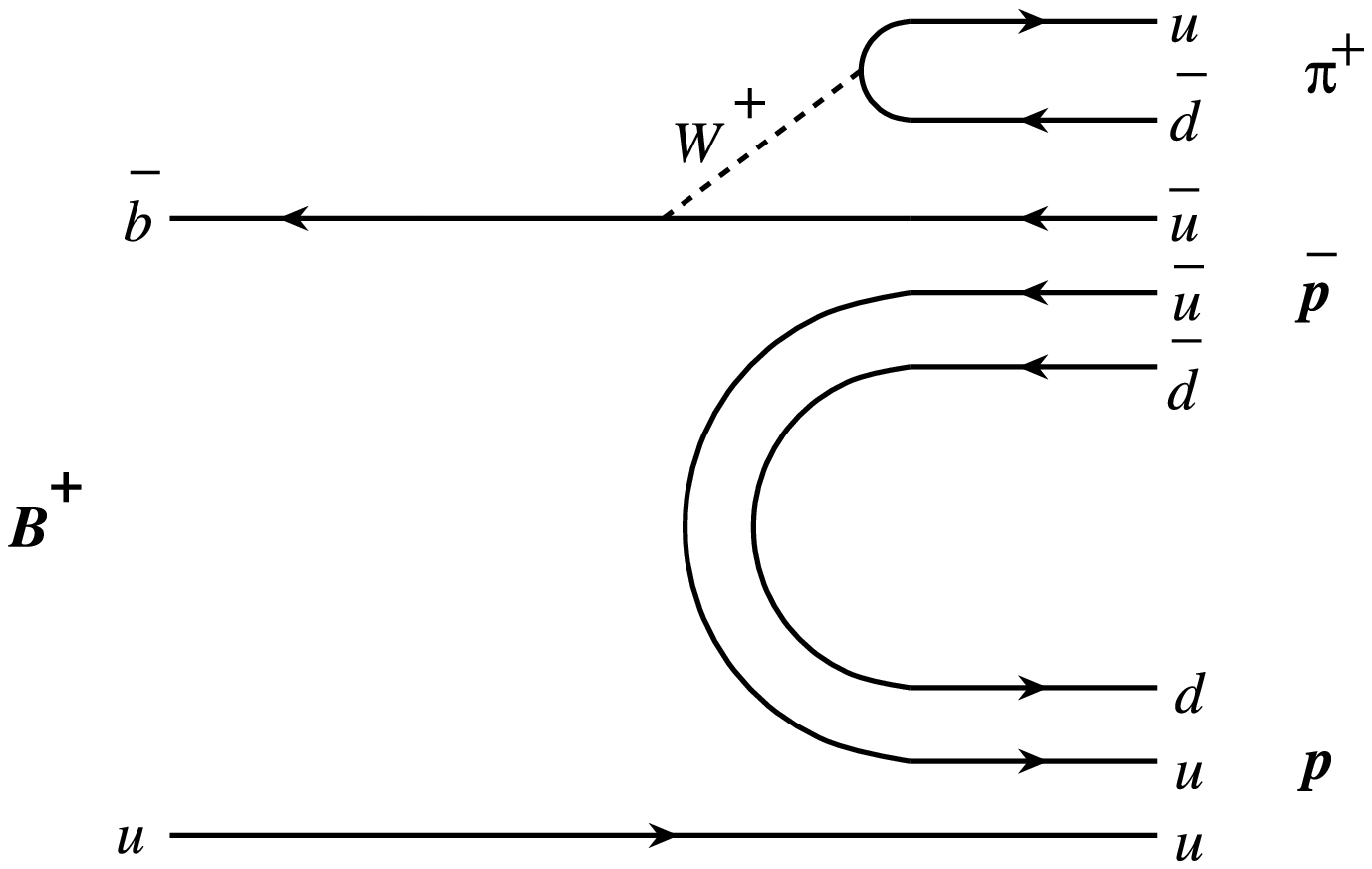}
\caption{ The possible leading (a) $b \to s$ penguin diagram
 and (b) $b \to u$ tree diagram 
for $\bp \to \ppk$ and $\bp \to \pppi$ decays, respectively. 
}
\label{fg:pppifeyn}
\end{center}
\end{figure}

The event selection criteria are based on information obtained
from the tracking system
(SVD and CDC) and the hadron identification system (CDC, ACC, and TOF).
All primary charged tracks
are required to satisfy track quality criteria
based on the track impact parameters relative to the
interaction point (IP). 
The deviations from the IP position are required to be within
$\pm$0.3 cm in the transverse ($x$--$y$) plane, and within $\pm$3 cm
in the $z$ direction, where the $z$ axis is opposite the
positron beam direction. For each track, the likelihood values $L_p$,
$L_K$, and $L_\pi$ that it is a proton, kaon, or pion, respectively,
are determined from the information provided by
the hadron identification system.  The track is identified as a proton
if $L_p/(L_p+L_K)> 0.6 $ and $L_p/(L_p+L_{\pi})> 0.6$, or as a kaon if
$L_K/(L_K+L_{\pi})> 0.6$, or as a pion if $L_{\pi}/(L_K+L_{\pi})> 0.6$.
For particles with
momenta at 2 GeV/$c$,
the proton selection efficiency is about 84\% (88\% for $p$ and 80\% for 
$\bar{p}$) and the fake rate is about
10\% for kaons and 3\% for pions; the kaon selection efficiency is about
85\% and the pion to kaon fake rate is about 2\%;
the pion selection efficiency is about 88\% and the kaon to pion 
fake rate is about 11\%.

Candidate $B$ mesons are reconstructed in the 
$\bp \to \ppk$ and $\bp \to \pppi$  modes.
We use two kinematic variables in the center of mass (CM) frame to identify the
reconstructed $B$ meson candidates: the beam energy
constrained mass $\mb = \sqrt{E^2_{\rm beam}-p^2_B}$, and the
energy difference $\de = E_B - E_{\rm beam}$, where $E_{\rm
beam}$ is the beam energy, and $p_B$ and $E_B$ are the momentum and
energy, respectively, of the reconstructed $B$ meson.
The candidate region is
defined as 5.20 GeV/$c^2 < \mb < 5.29$ GeV/$c^2$ and $-0.1$ GeV $ < \de< 0.3$
GeV. 
The lower bound in $\de$ for candidate events 
is chosen to exclude possible
cross-feed background from the decays with additional pions
to the search modes, e.g. $\bp \to \ppkst$.
From a GEANT~\cite{geant} based Monte Carlo (MC) simulation, the signal
peaks in a signal box defined by
5.27 GeV/$c^2 < \mb < 5.29$ GeV/$c^2$ and $|\de|< 0.05$ GeV,
and there is no peaking background except
cross-feed events between the $\ppk$ and $\pppi$ modes.

The background in the fit region arises dominantly from the continuum $e^+e^-
\to q\bar{q}$ ($q = u,\ d,\ s,\ c$) process.
We suppress the jet-like continuum background events relative to the more
spherical $B\bar{B}$ signal events using a Fisher discriminant~\cite{fisher}
that combines seven event shape variables, as described in Ref.~\cite{etapk}.
Probability density functions (PDFs) for the Fisher discriminant and
the cosine of the angle between the $B$ flight direction
and the beam direction in the $\Upsilon({\rm 4S})$ rest frame
are combined to form the signal (background)
likelihood ${\mathcal L}_{s}$ (${\mathcal L}_{b}$).
The signal PDFs are determined using signal MC
simulation; the background PDFs are obtained from 
the sideband data 
with $\mb < 5.26$ GeV/$c^2$.
We require
the likelihood ratio 
${\mathcal R} = {\mathcal L}_s/({\mathcal L}_s+{\mathcal L}_b)$ 
to be greater than 0.75 and 0.85 for the
$\ppk$ and $\pppi$ modes, respectively.
These selection
criteria are determined by optimization of $n_s/\sqrt{n_s+n_b}$, where $n_s$ 
and $n_b$
denote the expected numbers of signal and background events in the
signal box, respectively. 
We use the branching fractions from our 
previous measurements~\cite{polar,pph} in the 
calculation of $n_s$ and use the number of 
sideband events to estimate $n_b$. 
If there are  multiple $B$ candidates in a single event, we 
select the one with the best $\chi^2$ value from the 
vertex fit. 
The fractions of multiple $B$ events are about 8\% and 10\%
for the $\ppk$ and $\pppi$ modes, respectively.  

We perform an unbinned extended 
likelihood fit that maximizes the likelihood function, 
$$ L = {e^{-(N_s+N_b)} \over N!}\prod_{i=1}^{N} 
\left[\mathstrut^{\mathstrut}_{\mathstrut}N_s P_s(M_{{\rm bc}_i},\Delta{E}_i)+
N_b P_b(M_{{\rm bc}_i},\Delta{E}_i)\right],$$
to estimate the signal yield in the candidate region;
here $P_s\ (P_b)$ denotes the signal (background) PDF, 
$N$ is the number of events in the fit, and $N_s$ and $N_b$
are fit parameters representing the number of signal and background
events, respectively.

\vskip 1cm
\begin{figure}[htb]
\begin{center}
\hskip -10cm {\bf (a)}\\
\vskip -1.2cm
\includegraphics[width=0.6\textwidth]{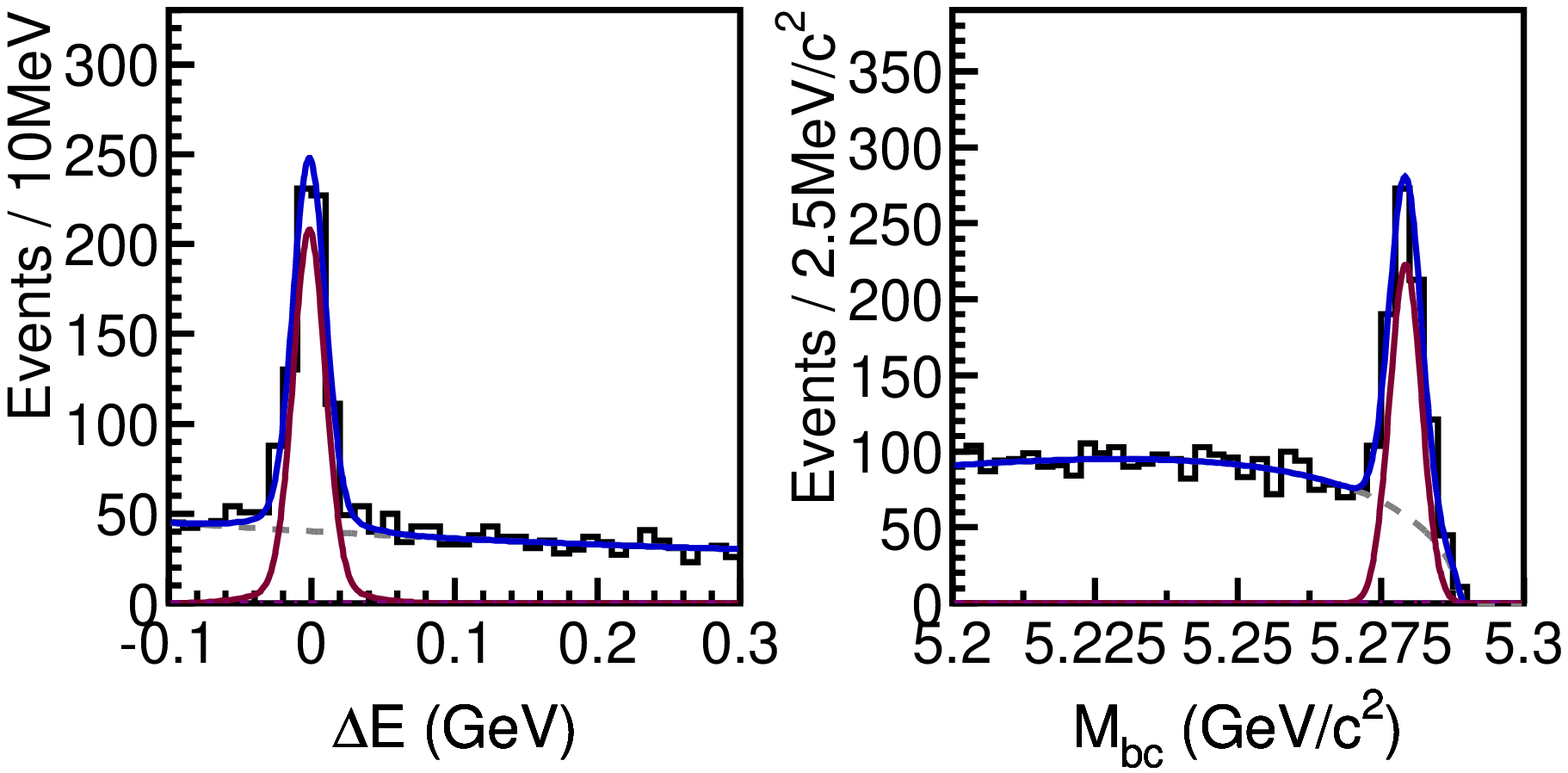}\\
\vskip 0.5cm
\hskip -10cm {\bf (b)}\\
\vskip -1.2cm
\includegraphics[width=0.6\textwidth]{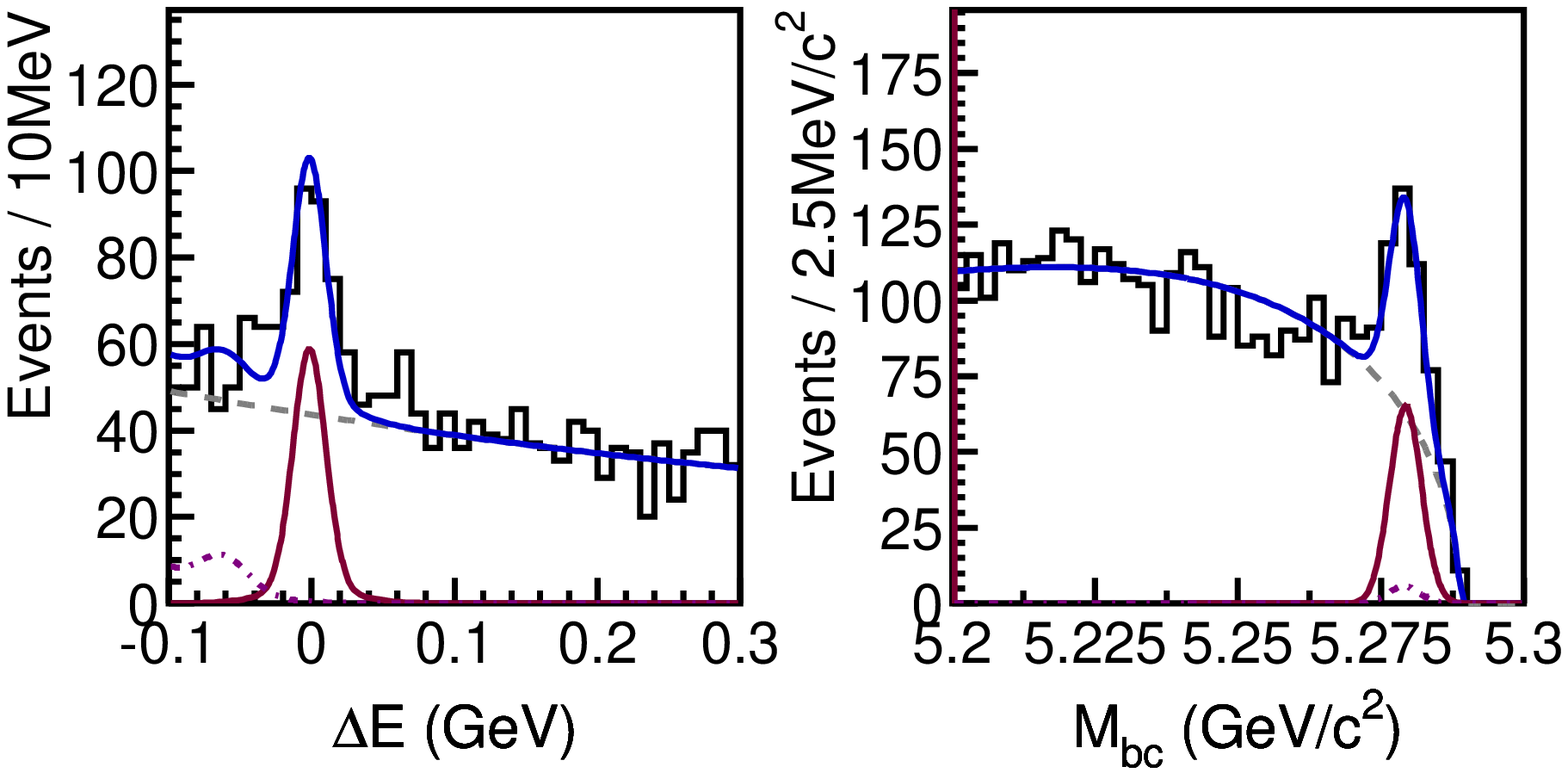}
\caption{ Distributions of $\de$ (with $\mb > 5.27$ GeV/$c^2$) and 
$\mb$ (with $|\de| < 0.05$ GeV), respectively, for
(a) $\ppk$ and (b) $\pppi$ modes with proton-antiproton pair mass less than  
2.85 GeV/$c^2$. 
The solid curves, solid peaks, and dashed curves represent the combined fit
result, fitted signal and fitted background, respectively. 
The dot-dashed curve indicates the $\ppk$ cross-feed background 
in the fit to the $\pppi$ mode.
}

\label{fg:mergembde}
\end{center}
\end{figure}

For the signal PDF, we use
a Gaussian function to represent the signal $\mb$
and a double Gaussian for $\de$
with parameters determined
by MC simulation. We then modify these parameters to account
for the discrepancies between data and MC using the $\ppk$ signal events ( 
$\mpp < 2.85$  GeV/$c^2$). With this correction, we can gain about 5\% more
signal yield.
The continuum background PDF 
is taken as the product of shapes in
$\mb$ and $\de$, which are assumed to be uncorrelated.
We use the parameterization first used by 
the ARGUS collaboration~\cite{Argus}, 
$ f(\mb)\propto \mb\sqrt{1-x^2}
\exp[-\xi (1-x^2)]$,  
to model
the $\mb$ background, with $x$ given by $\mb/E_{\rm beam}$ and $\xi$ as
a fit parameter. 
The $\de$ background shape is modeled by a normalized second order 
polynomial whose coefficients are fit parameters.
Because the $\pppi$ mode can contain
non-negligible cross-feed events from the $\ppk$ mode, we include the
$\ppk$ MC cross-feed shape in the fit for the determination of the
$\pppi$ yield. The cross-feed from $\pppi$ to $\ppk$ is negligible.
Figure~\ref{fg:mergembde} illustrates the fits of the $B$ yields
in a proton-antiproton mass region below 2.85 GeV/$c^2$,
which we refer to
as the threshold-mass-enhanced region.
The fitted $B$ yields are
632 $^{+29}_{-28}$ and
184 $^{+19}_{-19}$,
for the $\ppk$ and $\pppi$ modes, respectively.

\vskip 1.5cm
\begin{figure}[htb]
\begin{center}
\hskip 4.0cm {\bf (a)}\\
\vskip -2.1cm
\includegraphics[width=0.52\textwidth]{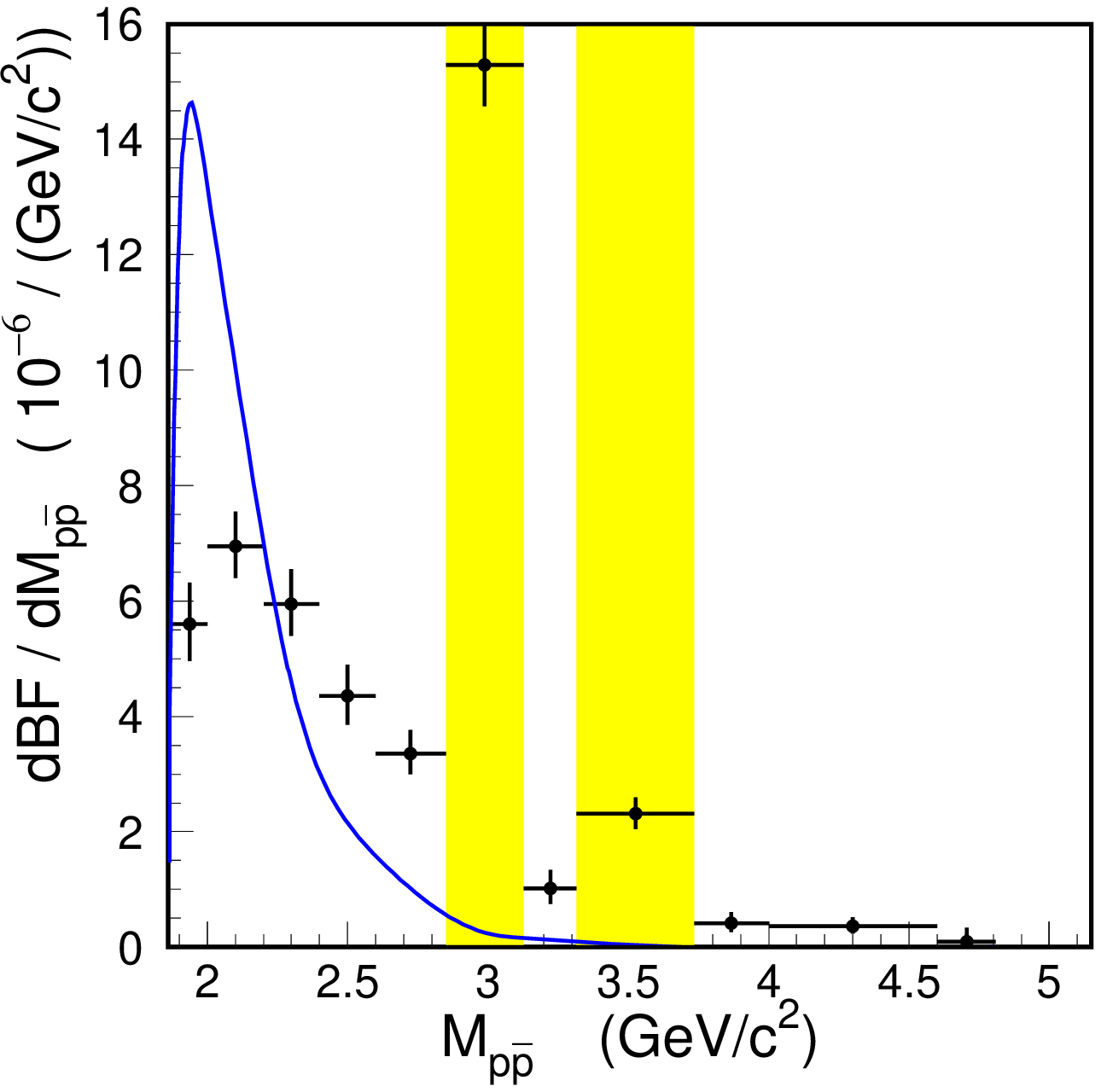}\\
\vskip 1cm
\hskip 4.0cm {\bf (b)}\\
\vskip -2.1cm
\includegraphics[width=0.52\textwidth]{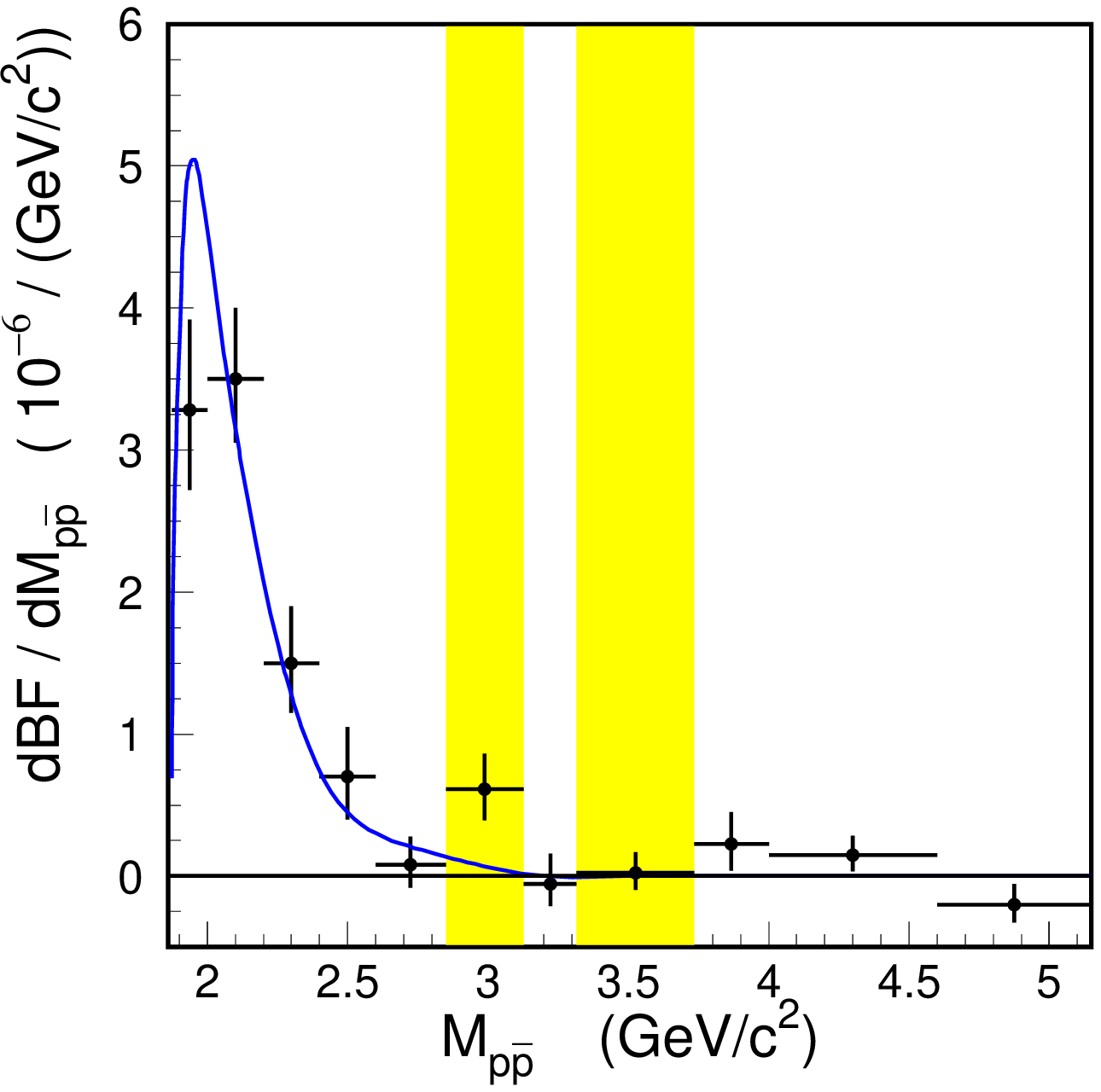}
\caption{Differential branching fractions for 
(a) $\ppk$ and (b) $\pppi$ 
modes 
as a function of proton-antiproton pair mass. 
The solid curves are theoretical predictions~[11] that are scaled to the 
observed charmless branching fractions.  
The two shaded mass bins,  $2.85<M_{p \bar{p}}<3.128$ GeV/$c^2$
and $3.315<M_{p \bar{p}}<3.735$ GeV/$c^2$, are not counted in 
the charmless
signal yields since they contain contributions
from the intermediate resonances $\eta_c, J/\psi$ and 
$\psi^{\prime},\chi_{c0},\chi_{c1}$ mesons, respectively.
}
\label{fg:allphase}
\end{center}
\end{figure}

Since there are two different
detector configurations and
the detection efficiency is dependent on $\mpp$,
we separate the data sample  into
two sets and determine the $B$ yields in bins of $\mpp$, 
where the signal PDF is assumed to be the same for all $\mpp$ bins.
We generate corresponding MC samples in order
to estimate the efficiencies properly. 
The partial branching fractions are obtained by correcting
the fitted B yields for the mass dependent efficiencies 
for each data set; they agree well with each other for the two data sets
and these results are
then combined to obtain the final results.

The differential branching fractions as a function of the 
proton-antiproton  mass for both $\ppk$ and $\pppi$ modes
are shown in Fig.~\ref{fg:allphase}, and the measured branching
fractions for different $\mpp$ bins are listed in 
Table~\ref{ppkresult-bins} and Table~\ref{pppiresult-bins}.  
Note that we have defined the charm veto: the regions
$2.850$ GeV/$c^2  <M_{p \bar{p}}<3.128$ GeV/$c^2$
and $3.315$ GeV/$c^2 <M_{p \bar{p}}<3.735$ GeV/$c^2$ are excluded
to remove background from $B$ decay modes
containing an $\eta_c$, $J/\psi$,
$\psi^{\prime}$, $\chi_{c0}$, or $\chi_{c1}$ meson.
These results supersede our previous measurements~\cite{pph,polar} 
with better accuracy.
The width of the $\pppi$ mode is narrower than that 
of the $\ppk$ mode and agrees better with the theoretical 
expectation~\cite{mass}. 
The error bars show the statistical uncertainties only.
The listed yield is the sum from the fits for two different periods;
the listed efficiency is
an effective one obtained by combining the two different detector configurations. 

Systematic uncertainties 
are determined using high-statistics control data samples. For proton
identification, we use a  $\Lambda \to p \pi^-$ sample, while for
$K/\pi$ identification we use a $D^{*+} \to D^0\pi^+$,
 $D^0 \to K^-\pi^+$ sample.
The average
efficiency difference for hadron identification
between data and MC has been corrected to obtain the final branching fraction
measurements. The corrections are about 9\% and 14\% for the $\ppk$
and $\pppi$ modes, respectively. The uncertainties associated with
the hadron identification corrections are estimated to be 4.2\% for two protons
and 1\% for one kaon/pion.
Tracking uncertainty is determined with
fully and partially reconstructed $D^*$ samples. It is about 1\% per 
charged track.
The $\mathcal R$ continuum suppression uncertainty is estimated from
 control samples with similar final states, 
 $\bp \to J/\psi K^+$  with $J/\psi \to \mu^+\mu^-$.
The uncertainties  
for $\mathcal R$ selection are 2.5\% and 4\% for the $\ppk$ and $\pppi$ modes,
respectively.
A systematic uncertainty of 2\% in the fit yield is determined by varying
the parameters of the signal and background PDFs.  
The MC statistical
uncertainty 
is less than 2\%. The error on the
number of $B\bar{B}$ pairs is 1.3\%, where we
assume  that the branching fractions of $\Upsilon({\rm 4S})$ 
to neutral and charged $B\bar{B}$ pairs are equal. 
The systematic uncertainties for each decay channel are 
summarized in Table~\ref{systematics}.
We first sum the correlated errors linearly and then combine them with the
uncorrelated ones in quadrature. The total systematic
uncertainties are 6.5\% and  7.4\% for 
the $\ppk$ and  $\pppi$ modes, respectively.

                                                                                

\begin{table}[b!]
\caption{The $B$ yields from $\de-\mb$ fits for the $\bp\to\ppk$ data sample,
detection efficiencies and
branching fractions (${\mathcal B}$) in different $\mpp$ regions.}
\label{ppkresult-bins}
\begin{center}
\begin{tabular}{c|ccc}
$\mpp$ (GeV/$c^2$)& \  Yield  & eff(\%) &${\mathcal B}$
($10^{-6}$)
\\
\hline $1.876-2.0$& \  $95.8^{+12.0}_{-11.0}$ & 30.6&
$0.70^{+0.09}_{-0.08} \pm 0.05$
\\
$2.0-2.2$& \  $188.0^{+16.1}_{-15.2}$& 30.0&
$1.39^{+0.12}_{-0.11} \pm 0.09$
\\
$2.2-2.4$& \  $146.1^{+14.1}_{-13.1}$& 27.3&
$1.19^{+0.12}_{-0.11} \pm 0.08$
\\
$2.4-2.6$& \  $99.9^{+12.2}_{-11.2}$& 25.5&
$0.87^{+0.11}_{-0.10} \pm 0.06$
\\
$2.6-2.85$& \  $100.7^{+12.0}_{-11.0}$& 26.6&
$0.84^{+0.10}_{-0.09} \pm 0.05$
\\
$2.85-3.128$& \  $496.8^{+23.9}_{-22.9}$& 26.0&
$4.25^{+0.20}_{-0.20} \pm 0.28$
\\
$3.128-3.315$& \  $20.9^{+6.8}_{-5.7}$& 25.2&
$0.19^{+0.06}_{-0.05} \pm 0.01$
\\
$3.315-3.735$& \  $108.2^{+13.1}_{-12.0}$& 24.8&
$0.97^{+0.12}_{-0.11} \pm 0.06$
\\
$3.735-4.0$& \  $11.7^{+5.9}_{-4.9}$& 24.3&
$0.11^{+0.05}_{-0.04} \pm0.01$
\\
$4.0-4.6$& \  $21.9^{+9.1}_{-7.9}$& 21.8&
$0.22^{+0.09}_{-0.08} \pm 0.01$
\\
$4.6-4.8$& \  $1.6^{+3.2}_{-2.1}$& 15.3&
$0.02^{+0.05}_{-0.03} \pm 0.00$
\\
\hline All  & \  & & $10.76^{+0.36}_{-0.33} \pm 0.70$
\\
\hline with charm veto  & \  &  & $5.54^{+0.27}_{-0.25} \pm 0.36$
\\
\hline $<2.85$ & \  &  & $5.00^{+0.24}_{-0.22} \pm 0.32$
\\
\end{tabular}
\end{center}
\end{table}

\begin{table}[b!]
\caption{The $B$ yields from $\de-\mb$ fits for the $\bp\to\pppi$ data sample,
detection efficiencies and
branching fractions (${\mathcal B}$) in different $\mpp$ regions.}
\label{pppiresult-bins}
\begin{center}
\begin{tabular}{c|ccc}
$\mpp$ (GeV/$c^2$)& \ Yield & eff(\%) &${\mathcal B}$
($10^{-6}$)
\\
\hline $1.876-2.0$& \  $51.3^{+9.5}_{-8.5}$ & 27.6&
$0.41^{+0.08}_{-0.07} \pm 0.03$
\\
$2.0-2.2$& \  $83.1^{+12.2}_{-11.2}$& 26.6&
$0.70^{+0.10}_{-0.09} \pm 0.05$
\\
$2.2-2.4$& \  $32.7^{+8.9}_{-7.8}$& 24.5&
$0.30^{+0.08}_{-0.07} \pm 0.02$
\\
$2.4-2.6$& \  $14.8^{+7.1}_{-5.9}$& 22.9&
$0.14^{+0.07}_{-0.06} \pm 0.01$
\\
$2.6-2.85$& \  $1.6^{+5.5}_{-4.3}$& 23.2&
$0.02^{+0.05}_{-0.04} \pm 0.00$
\\
$2.85-3.128$& \  $17.5^{+7.2}_{-6.1}$& 22.5&
$0.17^{+0.07}_{-0.06} \pm 0.01$
\\
$3.128-3.315$& \  $-0.5^{+4.3}_{-3.0}$& 22.3&
$-0.01^{+0.04}_{-0.03} \pm 0.00$
\\
$3.315-3.735$& \  $0.6^{+5.6}_{-4.4}$& 21.3&
$0.01^{+0.06}_{-0.05} \pm 0.00$
\\
$3.735-4.0$& \  $5.5^{+5.6}_{-4.3}$& 20.6&
$0.06^{+0.06}_{-0.05} \pm0.00$
\\
$4.0-4.6$& \  $8.2^{+7.9}_{-6.7}$& 20.8&
$0.09^{+0.08}_{-0.07} \pm 0.01$
\\
$4.6-5.15$& \  $-7.6^{+6.0}_{-4.7}$& 15.8&
$-0.11^{+0.08}_{-0.07} \pm 0.01$
\\
\hline All & \ & & $1.78^{+0.23}_{-0.19} \pm 0.13$
\\
\hline with charm veto  & \  &  & $1.60^{+0.22}_{-0.19} \pm 0.12$
\\
\hline $<2.85$& \ & & $1.57^{+0.17}_{-0.15} \pm 0.12$
\\
\end{tabular}
\end{center}
\end{table}

\vskip 1cm
\begin{figure}[htb]
\begin{center}
\hskip 3.5cm {\bf (a)}\\
\vskip -2.2cm
\includegraphics[width=0.54\textwidth]{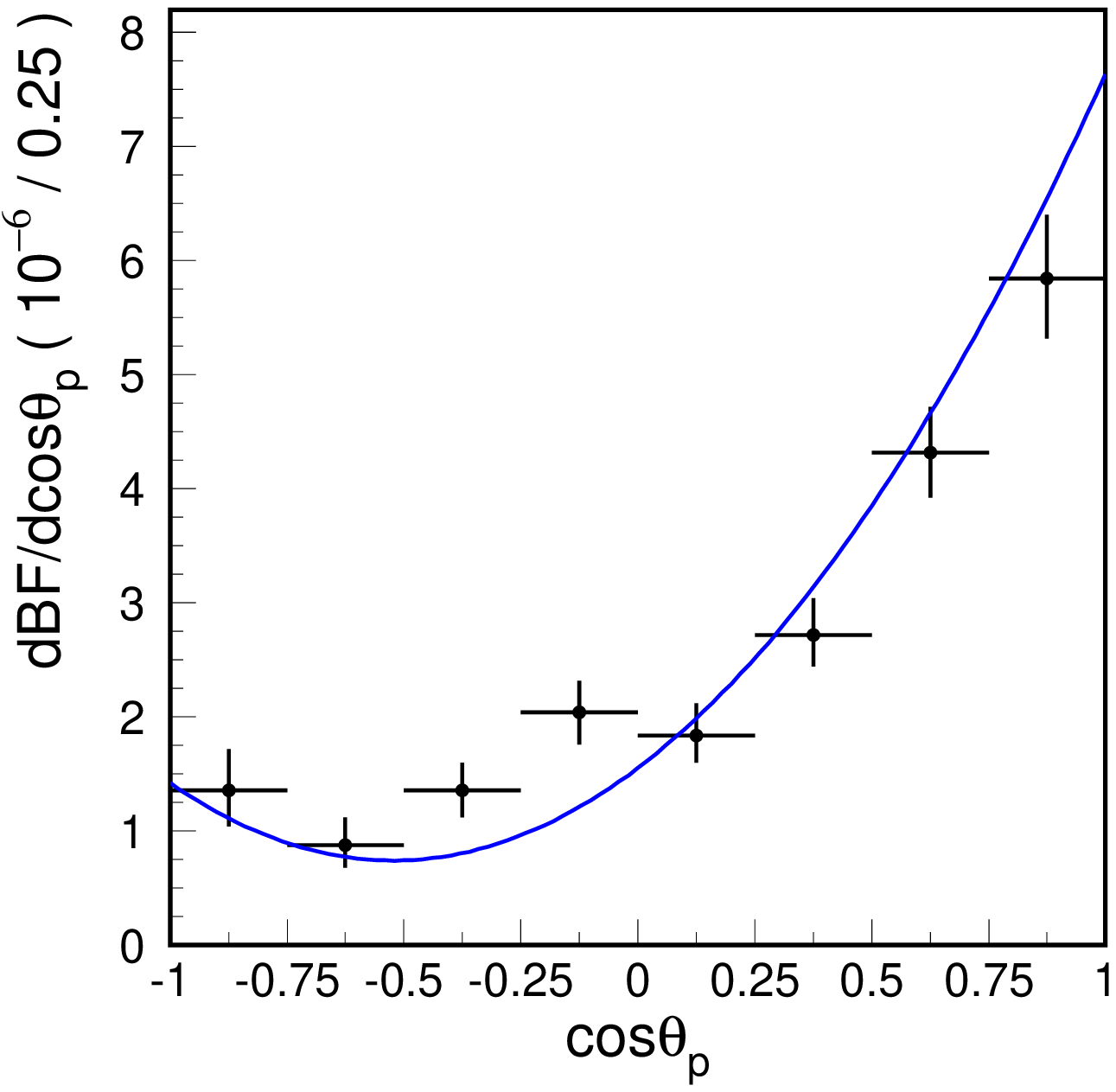}\\
\vskip 0.6cm
\hskip 3.5cm {\bf (b)}\\
\vskip -2.2cm
\includegraphics[width=0.54\textwidth]{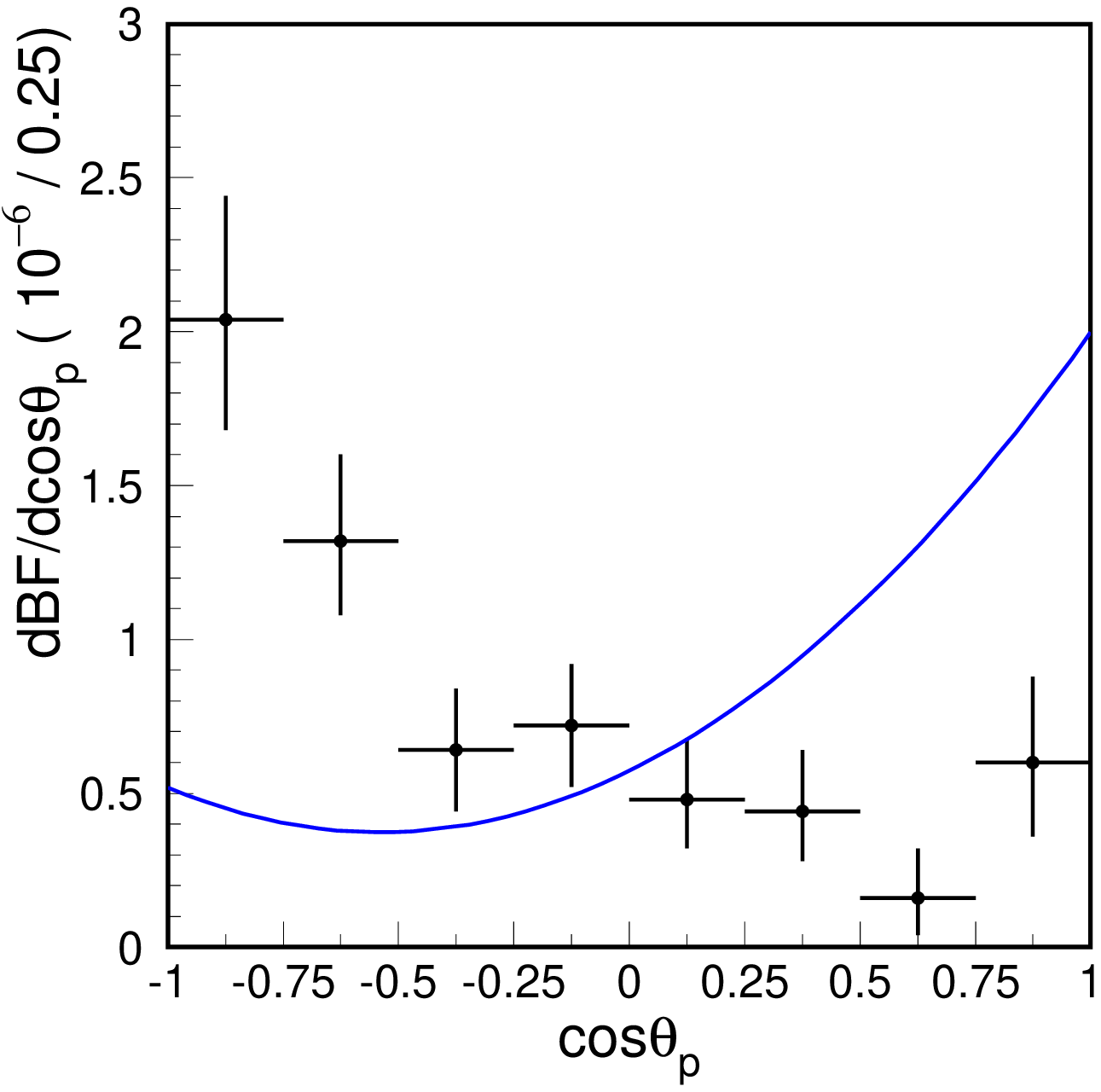}
\caption{Differential branching fractions {\it vs.} $\cos \theta_p$
 in the proton-antiproton pair system for (a) $\bp \to \ppk$ and 
(b) $\bp \to \pppi$. 
The solid curve is the theoretical prediction~[12].
}
\label{fg:pppicos}
\end{center}
\end{figure}

\begin{table}[htb]
\caption{Systematic uncertainties(\%) in the branching fraction for each decay
channel.}
\label{systematics}
\begin{center}
\begin{tabular}{c|cc}
Source & \  $\ppk$ & $\pppi$
\\
\hline
Tracking                        & \ $3.1$ & $3.2$

\\
Proton Identification           & \ $4.2$ & $4.2$
\\
K/$\pi$ Identification          & \ $1.0$ & $1.0$
\\
Likelihood Ratio Selection      & \ $2.5$ & $4.0$
\\
MC statistical error            & \ $1.4$ & $1.8$
\\
Fitting                         & \ $2.0$ & $2.0$
\\
Number of $B\bar{B}$ pairs      & \ $1.3$ & $1.3$
\\
\hline
Total                           & \ $6.5$ & $7.4$
\\
\\
\end{tabular}
\end{center}
\end{table}

We study the baryon angular distribution in the proton-antiproton
helicity frame at $M_\pp < 2.85$ GeV/$c^2$. 
The angle $\theta_p$ is defined as the
angle between the baryon direction and the oppositely charged meson 
direction 
in the 
proton-antiproton pair rest frame, i.e.   
this angle is determined 
by $p$ and $K^-$/$\pi^-$, or
by $\bar{p}$ and $K^+$/$\pi^+$.
We use the same likelihood method to estimate the $B$ yield 
in each $\theta_p$ bin. Again, the signal PDF is fixed and
the background shape is allowed to vary. 
The $\cos \theta_p$ distributions, shown in Fig.~\ref{fg:pppicos},  
for the $\ppk$ and $\pppi$ modes have opposite trends. 
This distribution for the $\pppi$  mode does not match the 
theoretical prediction~\cite{geng}, which is
based on an extrapolation of the $\ppk$ data
using the perturbative QCD framework. 
However, it does agree with the naive short distance
picture for a $b \to u$ 
weak decay. Particles directly associated with $b$ decay are more energetic
and the particle containing the spectator quark is generally less energetic.
After boosting to the proton-antiproton rest frame, the fast moving anti-protons
and $\pi^+$'s are back-to-back most of the time. 
However, the $b \to s \ {\rm 
gluon}$ process for the $\ppk$ case 
seems to completely disagree with
this short distance picture.  The baryon with the spectator
quark moves faster in the $B$ rest frame. The same phenomenon has been observed
in $\bz \to \plpi$~\cite{plh} decays. 
Another theoretical prediction proposes a 
long distance effect, namely 
$\pp$ rescattering through a hypothetical baryonium 
bound state~\cite{MSuzuki}, in order to explain the violation of 
the short distance picture for the $\ppk$ mode. Since this long distance
effect should also occur for the $\pppi$ case, it seems that
further theoretical investigations are needed to simultaneously explain
the behavior of both the $\ppk$ and $\pppi$ modes.

Because we have enough $\bp \to \ppk$ signal events in the threshold
enhancement region, we separate this region into five sub-regions. 
Fig.~\ref{fg:ppkcos}(a)-(e) shows the
efficiency corrected $B$ yield as a function of $\cos \theta_p$ for
these five sub-regions. 
We define the angular asymmetry as $A_{\theta_p} = {
{N_+ - N_-}\over
{N_+ + N_-}}$, where $N_+$ and
$N_-$
stand for the efficiency corrected $B$ yield with $\cos\theta_p > 0$ and
 $\cos\theta_p < 0$, respectively. The measured angular asymmetry 
as a function of $M_\pp$ is shown in Fig.~\ref{fg:ppkcos}(f).
It is interesting to see that there is a clear trend, which indicates
that the relative contributions from
two (or more) competing decay amplitudes are changing in this mass
range.  The measured average $A_{\theta_p}$ value of the threshold enhancement 
is given in Table~\ref{br-results}.
The systematic error, $\sim 0.03$, is determined
by checking the $\bp \to J/\psi K^+$ ($J/\psi \to
\mu^+\mu^-$) sample and the continuum background in $\bp \to \ppk$
where no asymmetry is expected. 
The observed $A_{\theta_p}$'s are $0.02 \pm 0.01$ for $\bp \to J/\psi K^+$
and $0.00 \pm 0.02$ for the continuum background.

\vskip 1cm
\begin{figure}[htb]
\begin{center}
\hskip -1.6cm {\bf (a)} \hskip 3.3cm {\bf (b)} \hskip 3.3 cm {\bf (c)}\\
\vskip -1.2cm
\includegraphics[width=0.28\textwidth]{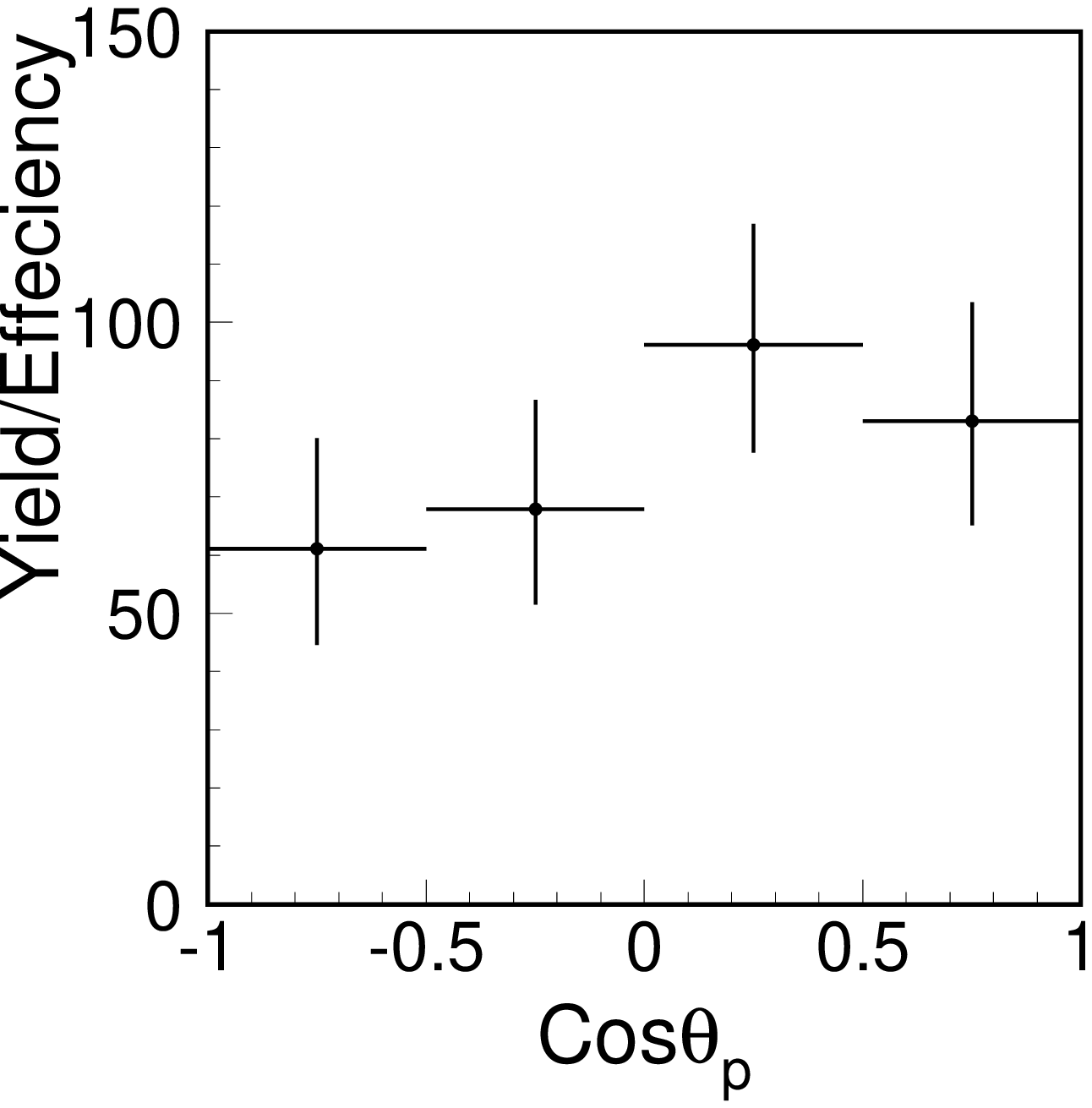}
\includegraphics[width=0.28\textwidth]{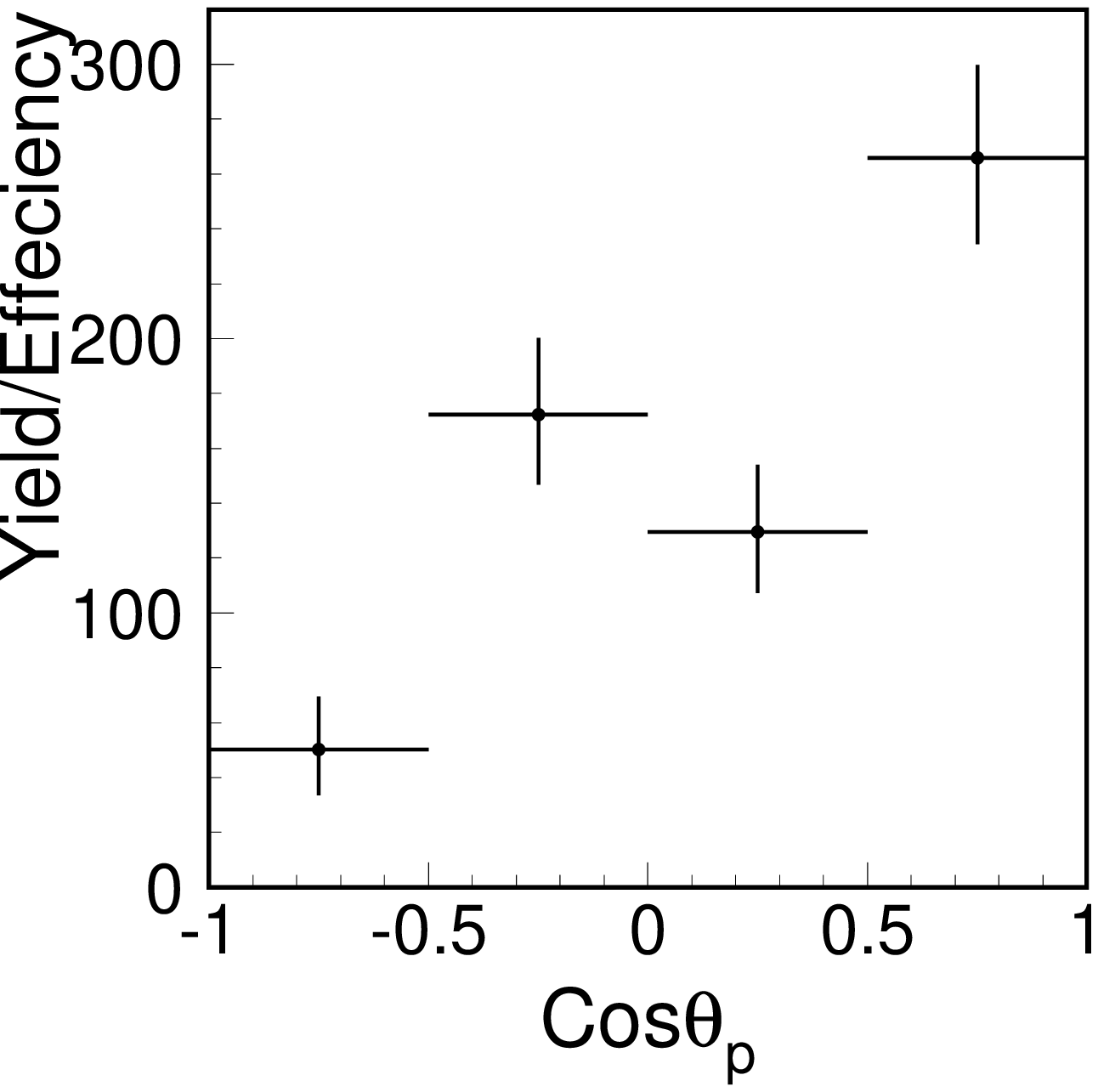}
\includegraphics[width=0.28\textwidth]{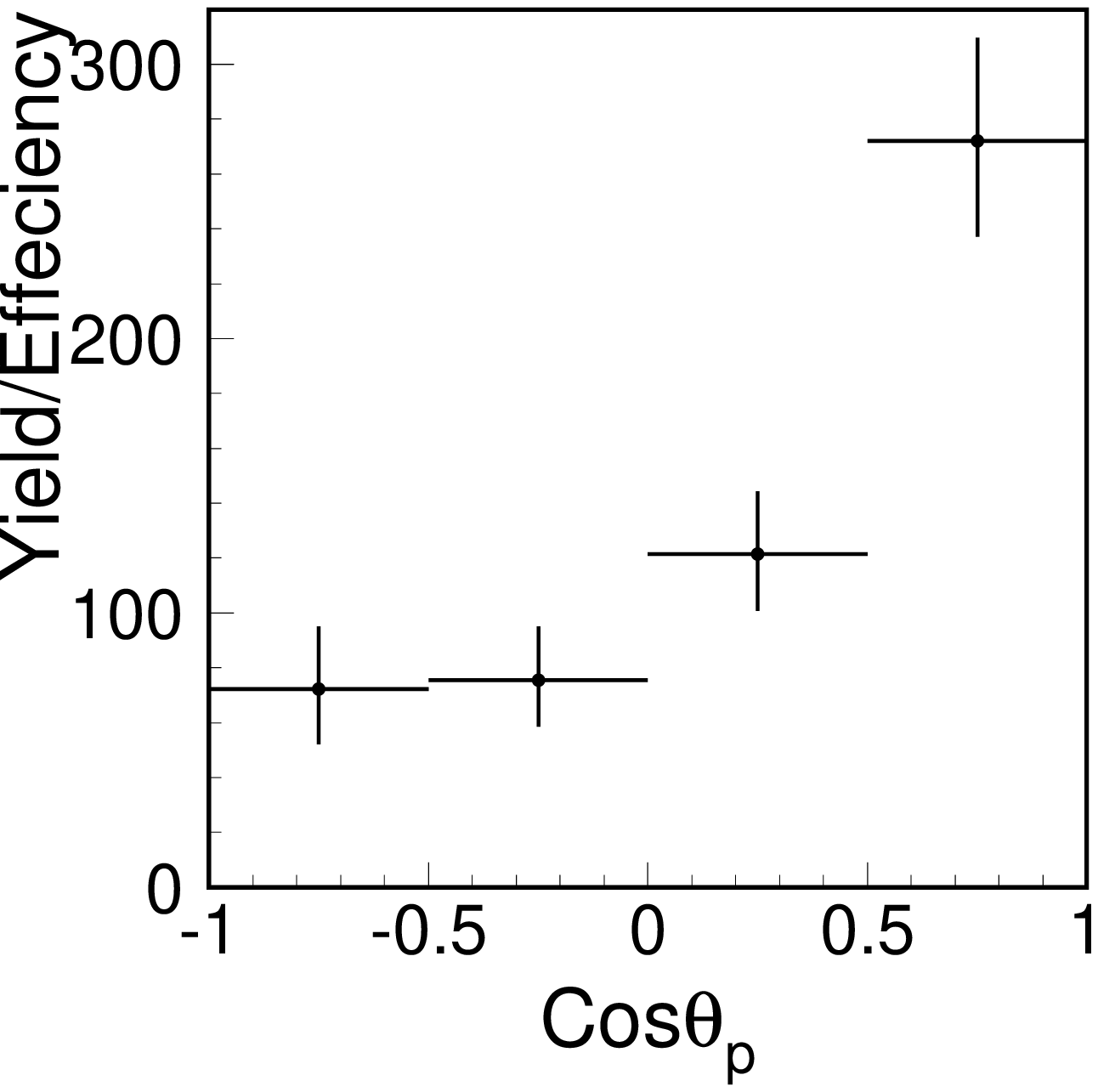}\\
\vskip 0.5cm
\hskip -1.6cm {\bf (d)} \hskip 3.3cm {\bf (e)} \hskip 3.3 cm {\bf (f)}\\
\vskip -1.2cm
\includegraphics[width=0.28\textwidth]{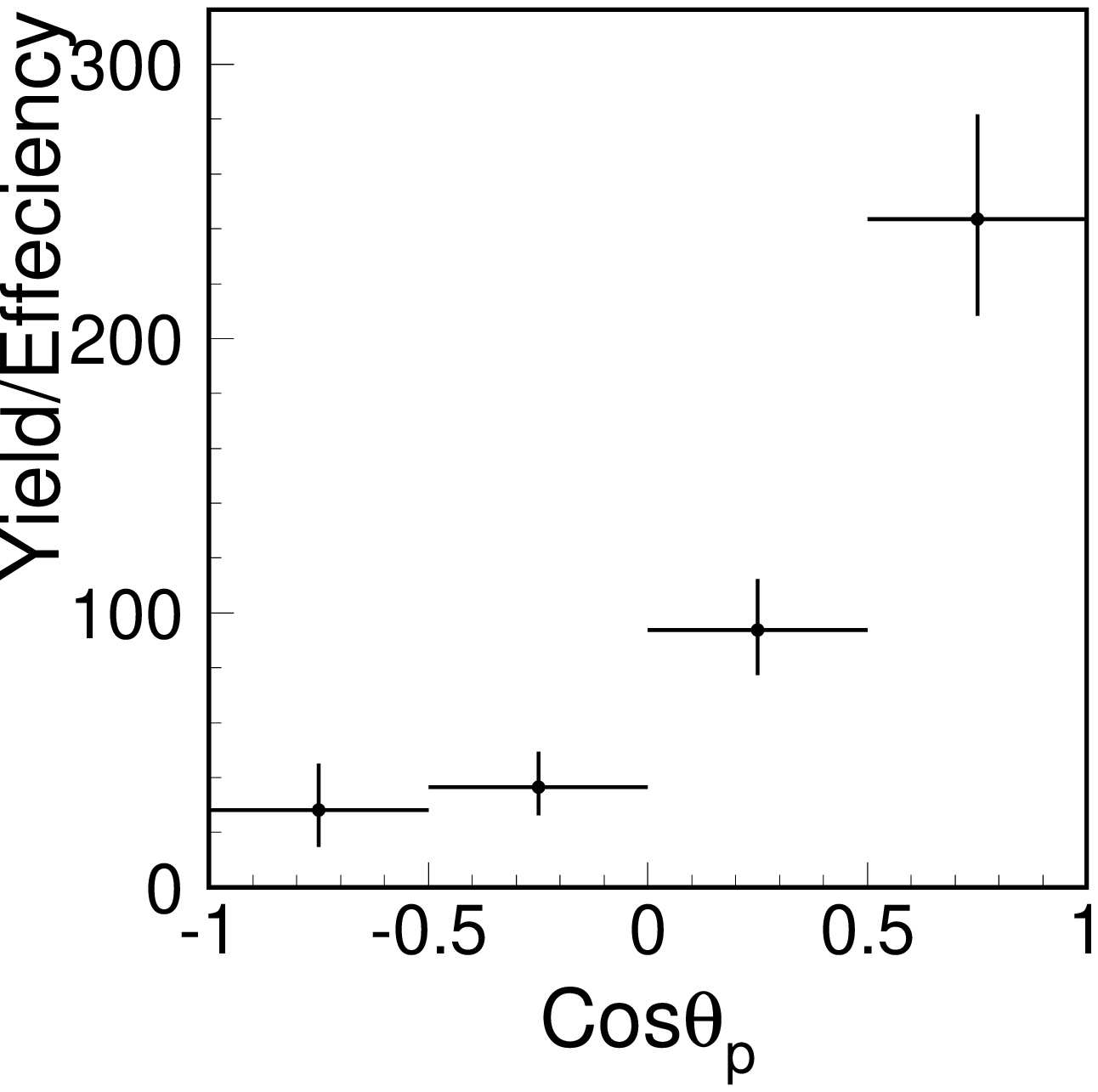}
\includegraphics[width=0.28\textwidth]{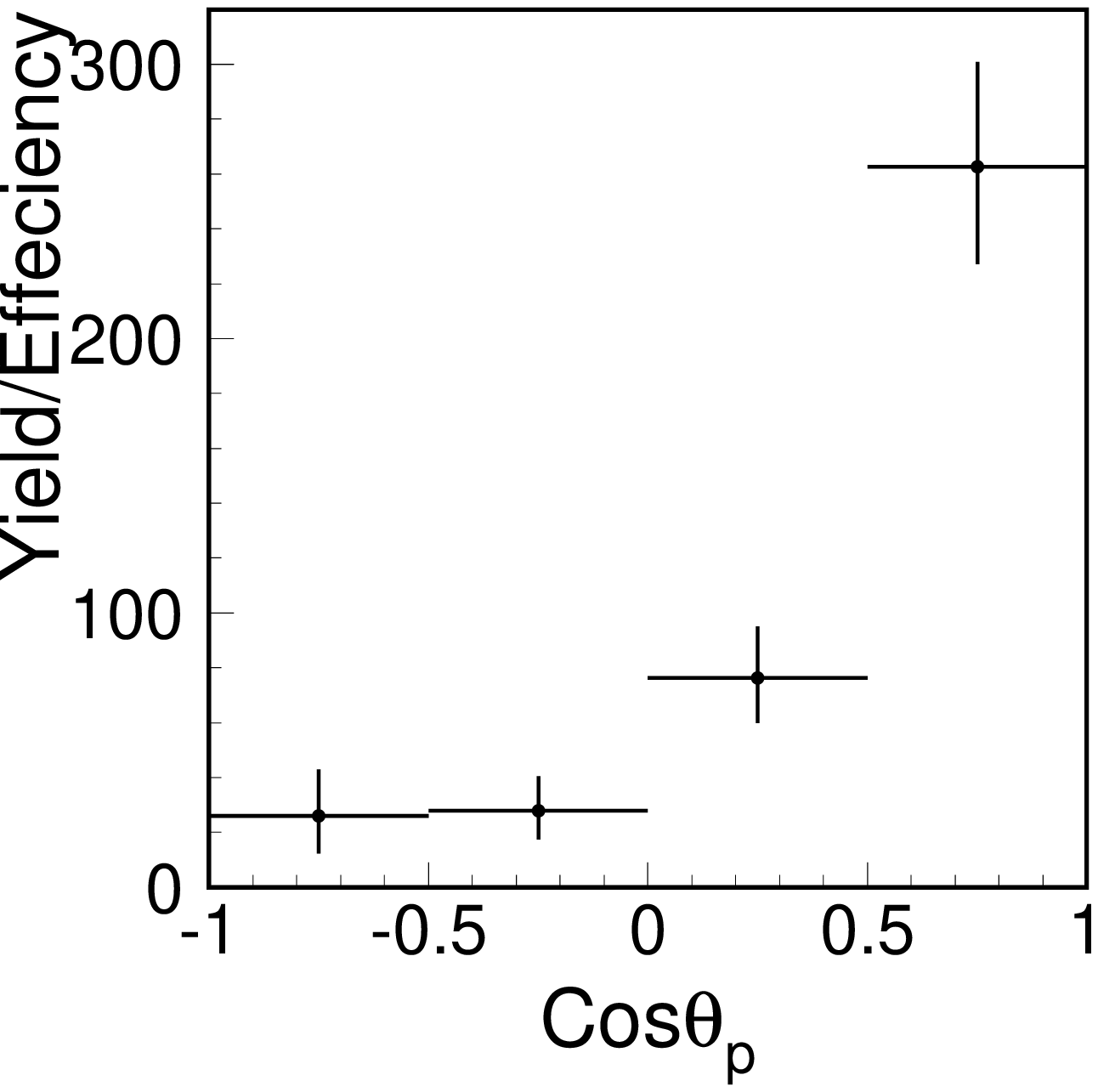}
\includegraphics[width=0.28\textwidth]{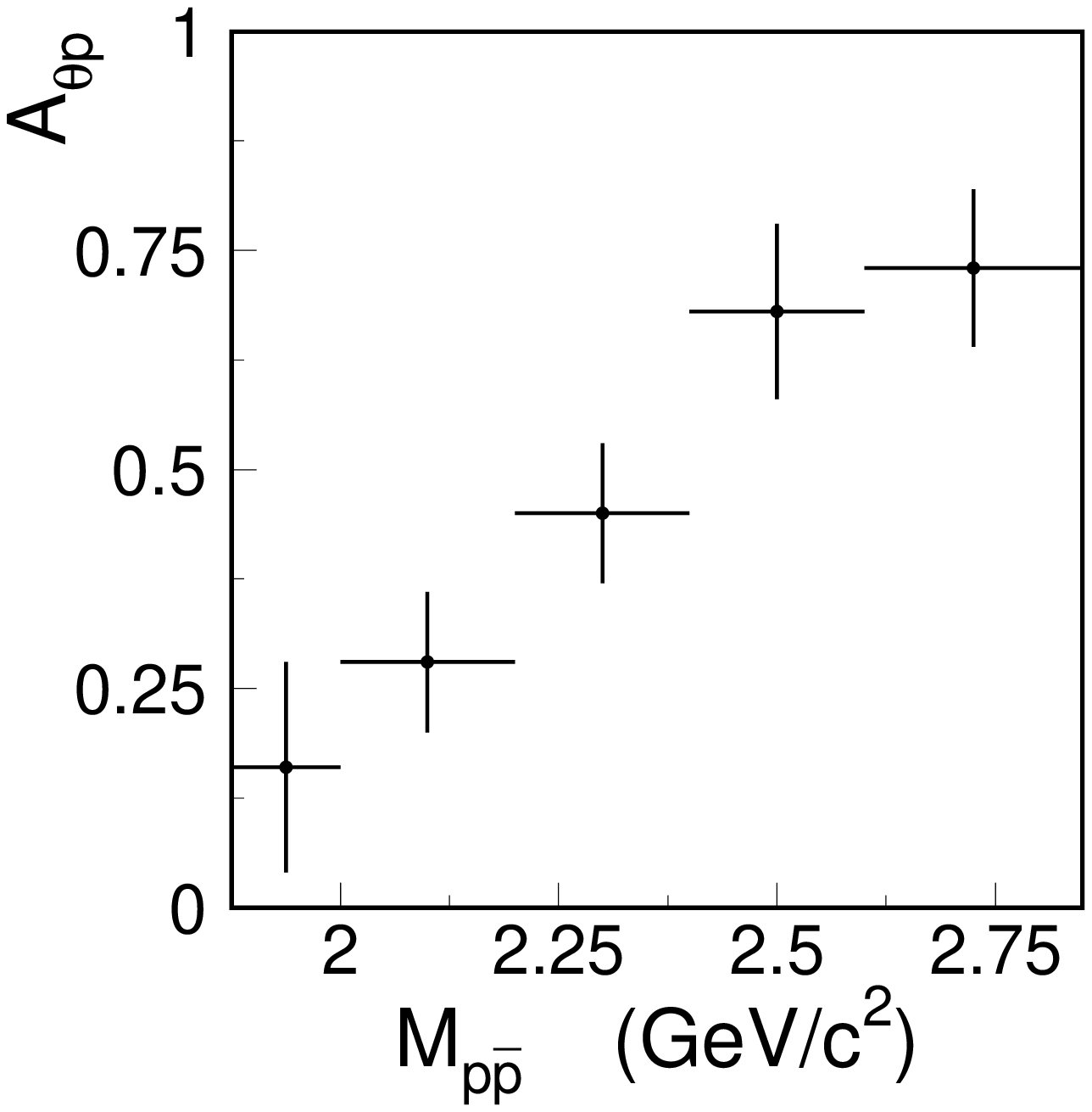}
\caption{ Efficiency corrected $B$ yield {\it vs.} $\cos \theta_p$ for
 (a) $M_{p \bar{p}} < 2.0$ GeV/$c^2$,
 (b) $2.0 < M_{p \bar{p}} < 2.2$ GeV/$c^2$, 
 (c) $2.2 < M_{p \bar{p}} < 2.4$ GeV/$c^2$, 
 (d) $2.4 < M_{p \bar{p}} < 2.6$ GeV/$c^2$, 
 and (e) $2.6 < M_{p \bar{p}} < 2.85$ GeV/$c^2$; 
(f) the measured angular 
asymmetries ($A_{\theta_p})$ for these five mass regions near threshold.
}
\label{fg:ppkcos}
\end{center}
\end{figure}

We also search for the
intermediate two-body decays, $\bp \to \bar{p} \Delta^{++}$  
($\Delta^{++} \to p \pi^+$)
and $\bp \to p {\bar\Delta^0}$ (${\bar\Delta^0} \to \bar{p}\pi^+$), 
from the $\pppi$ three-body final state. Events with  
$ M_{p\pi} < 1.4 $ GeV/$c^2$ are selected. 
No significant signals are found from the likelihood fit 
in those decay chains.
We observe $59$ and $86$ events in the signal box; 
the expected numbers of background events from the fits are $73.0 \pm 1.6$ and
 $81.4 \pm 1.6$ for  $\bp \to \bar{p} \Delta^{++}$ 
and  $\bp \to p {\bar\Delta^0}$, respectively.
We set upper limits
on the branching fractions at the 90\% confidence
level using the methods described in Refs.~\cite{Gary,Conrad} 
where the $7.4\%$ systematic uncertainty for $\bp \to \pppi$ is taken into
account. The  results are 
${\mathcal B}(\bp \to \bar{p} \Delta^{++}) < 0.14 \times 10^{-6}$  
and ${\mathcal B}(\bp \to p {\bar\Delta^0}) <1.38 \times 10^{-6}$. 
These numbers are smaller than the theoretical expectations 
but agree with other experimental findings~\cite{Tsai}.

Since there is a prediction~\cite{jcp} that direct $CP$ violation 
in $\bp \to J/\psi K^+$ is at the 1\% level, it is quite possible that this
effect could be magnified due to the interference~\cite{etaccp}
between the resonance and the threshold enhancement.  
We define the charge asymmetry $A_{CP}$
as $(N_{b} - N_{\bar{b}})/ (N_{b} + N_{\bar{b}})$ for the $\ppk$ 
and $\pppi$ modes,
where 
$N_b (N_{\bar{b}})$ stands for
the efficiency corrected $B^-$  ($B^+$) yield. 
The selection criteria for J/$\psi$ ($\eta_c$) and related consistency checks
have been reported in Ref.~\cite{Wu}. We adopt the same criteria and assume
the signal PDFs are the same for  both $B^-$ and $B^+$ samples.
The results from the likelihood fits
are listed in Table~\ref{br-results} for various mass/resonance regions. 
No significant
charge asymmetries are found.
The systematic uncertainty is assigned using 
the measured charge asymmetry for sideband data and is found to be 
$-0.01 \pm 0.01$.



In summary, using 449 $ \times 10^6 B\bar{B}$ events, we measure the
 mass and the angular distributions of the proton-antiproton pair
system near threshold for the $\ppk$ and  $\pppi$  
baryonic $B$ decay modes. 
These results supersede our previous measurements~\cite{pph,polar} 
with better accuracy.
The width of the threshold enhancement in the
$\pppi$ mode is narrower than that 
of the $\ppk$ mode and agrees better with the theoretical 
expectation~\cite{mass}. 
The proton polar angular distributions of the $\ppk$ and $\pppi$ modes have 
opposite trends. This shows that the $b \to s$ and $b \to u$ processes 
are kinematically different at short distance.
We also search for intermediate two-body decays in the 
$\pppi$ final states; no significant
signals are found. 

\begin{table}[htb]
\caption{Summary of the results in the mass region $M_\pp < 2.85$
GeV/$c^2$ and in other intermediate resonance regions.
Y is the fitted signal yield (or upper limit at 90\% confidence level),
 {$\mathcal{B}$} is the branching
fraction,
$A_{\theta}$ is the angular asymmetry and
$A_{CP}$ is the charge asymmetry.}
\label{br-results}
\begin{center}
{
\begin{tabular}{c|cccc}
Mode &\  Y & {$\mathcal B$} ($10^{-6}$)    &$A_{\theta}$ & $A_{CP}$
\\
\hline $\ppk$ & \ $632^{+29}_{-28}$ & $5.00^{+0.24}_{-0.22}\pm
0.32$ & $0.45 \pm 0.05 \pm 0.03$ & $-0.02 \pm 0.05 \pm 0.02$
\\
     $\eta_c K^+$ & \ $ 158^{+14}_{-13} $ & - 
& - & $-0.16 \pm 0.08 \pm 0.02$
\\
     $J/\psi K^+$ & \ $236^{+16}_{-16}$ & - 
& - & \ \ $0.09 \pm 0.07 \pm 0.02$
\\
\hline $\pppi$ & \ $184^{+19}_{-19}$ & $1.57^{+0.17}_{-0.15}\pm
0.12$ & $-0.47 \pm 0.12 \pm 0.03$ 
& $-0.17 \pm 0.10 \pm 0.02$
\\
       $\bar{p}\Delta^{++}$ & \ $<7.5$ &$<0.14$ & - & -
\\
      $p{\bar\Delta^0}$ & \ $<25.9$& $<1.38$ & - & -
\\
\end{tabular}
}
\end{center}
\end{table}


We thank the KEKB group for the excellent operation of the
accelerator, the KEK cryogenics group for the efficient
operation of the solenoid, and the KEK computer group and
the National Institute of Informatics for valuable computing
and Super-SINET network support. We acknowledge support from
the Ministry of Education, Culture, Sports, Science, and
Technology of Japan and the Japan Society for the Promotion
of Science; the Australian Research Council and the
Australian Department of Education, Science and Training;
the National Science Foundation of China and the Knowledge
Innovation Program of the Chinese Academy of Sciences under
contract No.~10575109 and IHEP-U-503; the Department of
Science and Technology of India;
the BK21 program of the Ministry of Education of Korea,
the CHEP SRC program and Basic Research program
(grant No.~R01-2005-000-10089-0) of the Korea Science and
Engineering Foundation, and the Pure Basic Research Group
program of the Korea Research Foundation;
the Polish State Committee for Scientific Research;
the Ministry of Education and Science of the Russian
Federation and the Russian Federal Agency for Atomic Energy;
the Slovenian Research Agency;  the Swiss
National Science Foundation; the National Science Council
and the Ministry of Education of Taiwan; and the U.S.\
Department of Energy.


\begin{thebibliography}{99}
\bibitem{conjugate} {Throughout this report,
inclusion of charge conjugate mode is always implied unless
otherwise stated.}
\end{thebibliography}
\end{document}